%% file: main.tex
\documentclass[letterpaper]{article} 
\usepackage[preprint]{aaai2027}
\usepackage[hyphens]{url}  
\usepackage{graphicx} 
\usepackage{natbib}  
\usepackage{caption} 
\usepackage{algorithm}
\usepackage{algorithmic}

\usepackage{newfloat}
\usepackage{listings}
\DeclareCaptionStyle{ruled}{labelfont=normalfont,labelsep=colon,strut=off} 
\floatstyle{ruled}
\newfloat{listing}{tb}{lst}{}
\floatname{listing}{Listing}

\usepackage{booktabs}
\usepackage{colortbl}
\usepackage{multirow}
\usepackage{tabularx}
\usepackage{xspace}
\definecolor{modelgray}{RGB}{235,235,235}
\definecolor{oursblue}{RGB}{224,242,249}

\newcommand{\sweprime}{\textsc{SWE-Prime}\xspace}

\title{SWE-Prime: Fewer Trajectories, Better Performance}
\author{
    Dewu Zheng\textsuperscript{\rm 1}\equalcontrib,
    Ruizhe Ye\textsuperscript{\rm 2}\equalcontrib,
    Yanlin Wang\textsuperscript{\rm 1}\corresponding,
    Yang Ye\textsuperscript{\rm 2},
    Hongyu Zhang\textsuperscript{\rm 3},\\
    Ensheng Shi\textsuperscript{\rm 2},
    Xilin Liu\textsuperscript{\rm 2},
    Yuchi Ma\textsuperscript{\rm 2},
    Jianxing Yu\textsuperscript{\rm 1},
    Zibin Zheng\textsuperscript{\rm 1}
}
\affiliations{
    \textsuperscript{\rm 1}Sun Yat-sen University, China\\
    \textsuperscript{\rm 2}Huawei Cloud Computing Technologies Co., Ltd., China\\
    \textsuperscript{\rm 3}Chongqing University, China\\
    zhengdw5@mail2.sysu.edu.cn, \{wangylin36, yujx26, zhibin\}@mail.sysu.edu.cn\\
    \{yeruizhe, yeyang14, shiensheng, liuxilin3, mayuchi1\}@huawei.com\\
    hyzhang@cqu.edu.cn
}

\begin{document}

\maketitle
\input{abstract.tex}

\input{body.tex}

\bibliography{ref.bib}


\end{document}

%% file: abstract.tex
\begin{abstract}
To improve large language models' ability to resolve real-world software issues, prior work has focused on constructing large-scale agent trajectory datasets and performing supervised fine-tuning (SFT) on successful trajectories. However, task success does not guarantee high-quality supervision: successful trajectories may still contain ineffective, redundant, or risky steps. Directly using such trajectories for SFT can introduce noisy supervision and encourage models to imitate undesirable problem-solving behaviors. Therefore, we propose \sweprime, a multi-granularity, two-stage SFT data selection method that progressively filters training data at the trajectory and segment levels. Specifically, the first stage performs trajectory-level screening based on process quality, result quality, and data representativeness, selecting a high-quality and representative subset of successful trajectories. The second stage performs segment-level selection by grouping consecutive steps into semantic segments and assessing each segment based on its contribution to the final solution, learnability, and potential risks. During SFT, all segments remain in the sequence to preserve context, while only selected segments contribute to the loss computation. Experiments on SWE-Bench Pro and SWE-Bench Verified show that training on the 10\% trajectory subset selected by \sweprime outperforms training on the full resolved dataset, yielding relative performance gains of up to 12.2\% and 24.2\%, respectively.
\end{abstract}

%% file: body.tex
\section{Introduction}

Software issue resolution in real-world repositories has received growing attention in recent years~\cite{jimenez2024swebench,pan2024swegym,yang2025swesmith,li2026issueresolutionsurvey}. To improve coding agents' performance on this long-horizon task, recent work has constructed large-scale software engineering datasets for supervised fine-tuning (SFT)~\cite{pan2024swegym,yang2025swesmith,jain2025r2egym,guo2025swefactory,zheng2023codellmsurvey}. For example, SWE-Gym derives executable tasks from real-world GitHub issues, while SWE-smith and R2E-Gym use synthetic data generation to scale task and trajectory collection. These approaches then perform SFT on successful trajectories identified through execution-based verification.

\begin{figure*}[t]
    \centering
    \includegraphics[width=\textwidth]{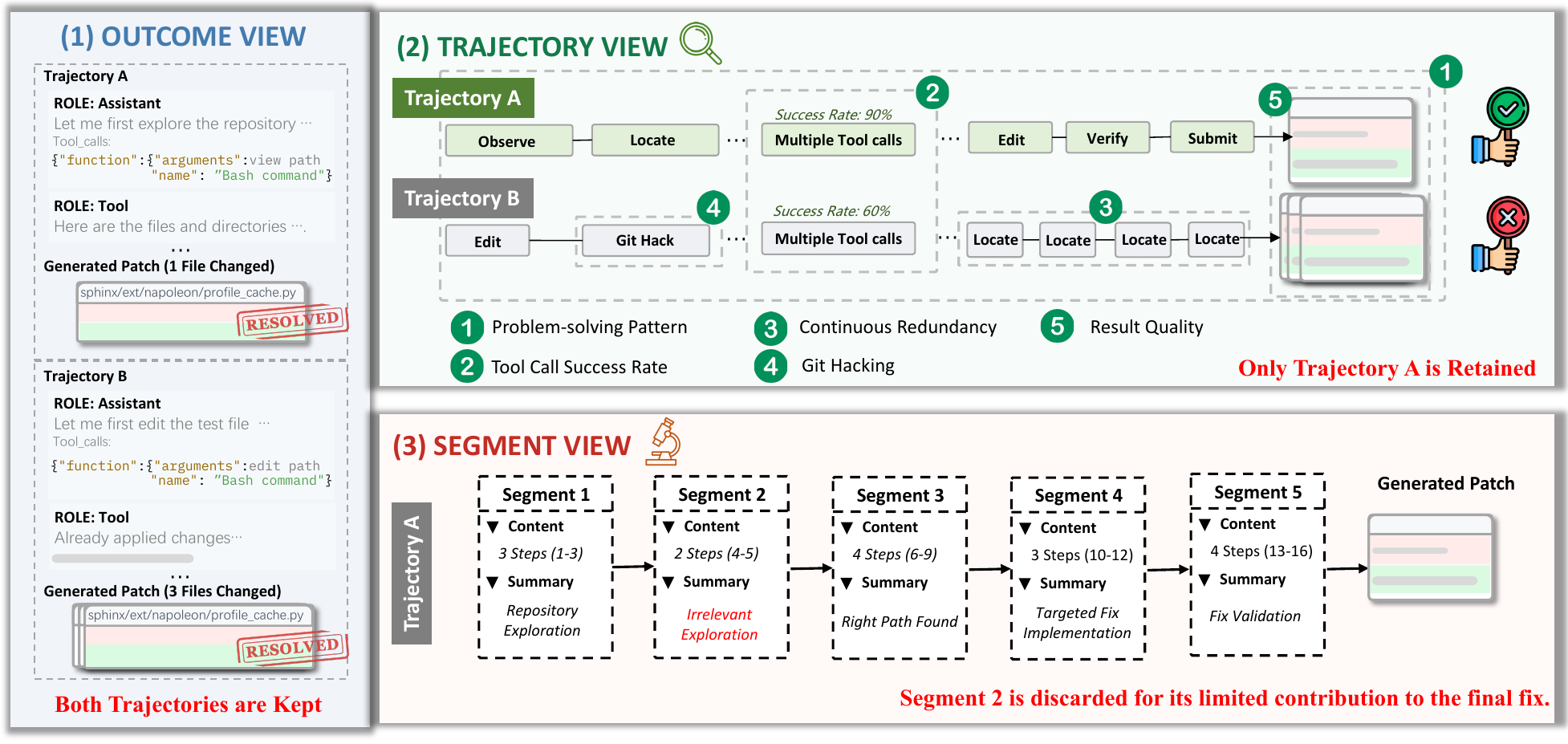}
    \caption{Motivating example comparing outcome-only, trajectory-level, and segment-level data selection.}
    \label{fig:motivating-example}
\end{figure*}

However, task success alone does not guarantee high-quality supervision~\cite{uesato2022processfeedback,lightman2023verify,xiong2024watch,wang2026solved}. Successful trajectories may still contain ineffective, redundant, or risky behaviors that should not be used as supervision for SFT~\cite{chen2025atlas,chen2025step,wang2025steca}. Figure~\ref{fig:motivating-example} illustrates this issue from three perspectives.

\noindent\textbf{Outcome View.} Existing work adopts an outcome-based view and thus retains both successful trajectories~\cite{pan2024swegym,yang2025swesmith,jain2025r2egym,guo2025swefactory}.

\noindent\textbf{Trajectory View.} The trajectory view shows why Trajectory B is unsuitable for SFT through the following quality aspects:

\begin{enumerate}
    \renewcommand{\labelenumi}{\textbf{(\arabic{enumi})}}
    \item Problem-solving Pattern: Standard problem-solving processes help models learn systematic issue-resolution behaviors during SFT~\cite{chen2026beyondfinalcode}. Trajectory A follows an observe-edit-verify workflow, whereas Trajectory B modifies the code without prior exploration.
    \item Tool Call Success Rate: Frequent tool failures introduce unreliable tool-use patterns into SFT supervision~\cite{yang2025toolmind}. Trajectory B contains more such failures, whereas Trajectory A demonstrates stable tool use.
    \item Continuous Redundancy: Long-context repetition caused by losing track of prior actions introduces undesirable SFT supervision~\cite{xiao2025agentdiet}. Accordingly, Trajectory B should be excluded because it repeats the same localization action without obtaining new information.
    \item Git Hacking: Leakage-based shortcuts introduce cheating behaviors into SFT supervision~\cite{song2026swemaster,tao2026swelego}. Trajectory B is unsuitable for SFT because it retrieves target repair information from version history.
    \item Result Quality: SFT should favor minimal yet sufficient patches to reduce regression risks~\cite{chen2025patchquality}. Trajectory A follows this principle, whereas Trajectory B modifies additional unnecessary files.
\end{enumerate}

\noindent\textbf{Segment View.} Even a trajectory assessed as high-quality at the trajectory level may still contain segments of limited value for SFT~\cite{xiong2024watch,wang2025steca,deng2025agentpro}. Segment 2 exemplifies such low-value behavior. Its steps execute successfully, but the segment does not contribute to the final fix, making it unsuitable for supervision. We assess semantic segments rather than individual steps because a problem-solving behavior often spans multiple actions and observations that should be evaluated together to determine its intent, outcome, and contribution.

These observations suggest that high-quality SFT data should be compact and low-noise, providing demonstrations that models can reliably learn from~\cite{chen2024alpagasus,liu2024deita}. \textbf{To this end, we propose \sweprime, a multi-granularity, two-stage data selection method that progressively filters training data from the trajectory level to the semantic segment level.} 
Specifically, Stage~1 operates at the trajectory level to select a high-quality and representative subset of successful trajectories. Stage~2 operates at the semantic segment level to identify high-value behaviors within the retained trajectories. 
During SFT, all segments remain in the sequence to preserve context, while only selected segments contribute to the loss. 

We evaluate \sweprime on two representative benchmarks, SWE-Bench Pro and SWE-Bench Verified. Models trained on the 10\% subset selected by \sweprime achieve relative gains of up to 12.2\% and 24.2\% over models trained on all successful trajectories, respectively.

Our main contributions are summarized as follows:

\begin{itemize}
    \item We highlight that prior work largely overlooks differences in data quality among successful trajectories.
    \item We introduce semantic segments as behaviorally coherent units for fine-grained supervision-quality assessment in long-horizon trajectories, rather than isolated steps.
    \item We propose \sweprime, a multi-granularity, two-stage method designed to provide compact, low-noise, and informative supervision for coding-agent SFT.
    \item Through extensive experiments across three models and two benchmarks, we demonstrate that \sweprime substantially outperforms SFT on the full dataset while using only 10\% of the trajectories.
\end{itemize}

\begin{figure*}[t]
    \centering
    \includegraphics[width=\textwidth]{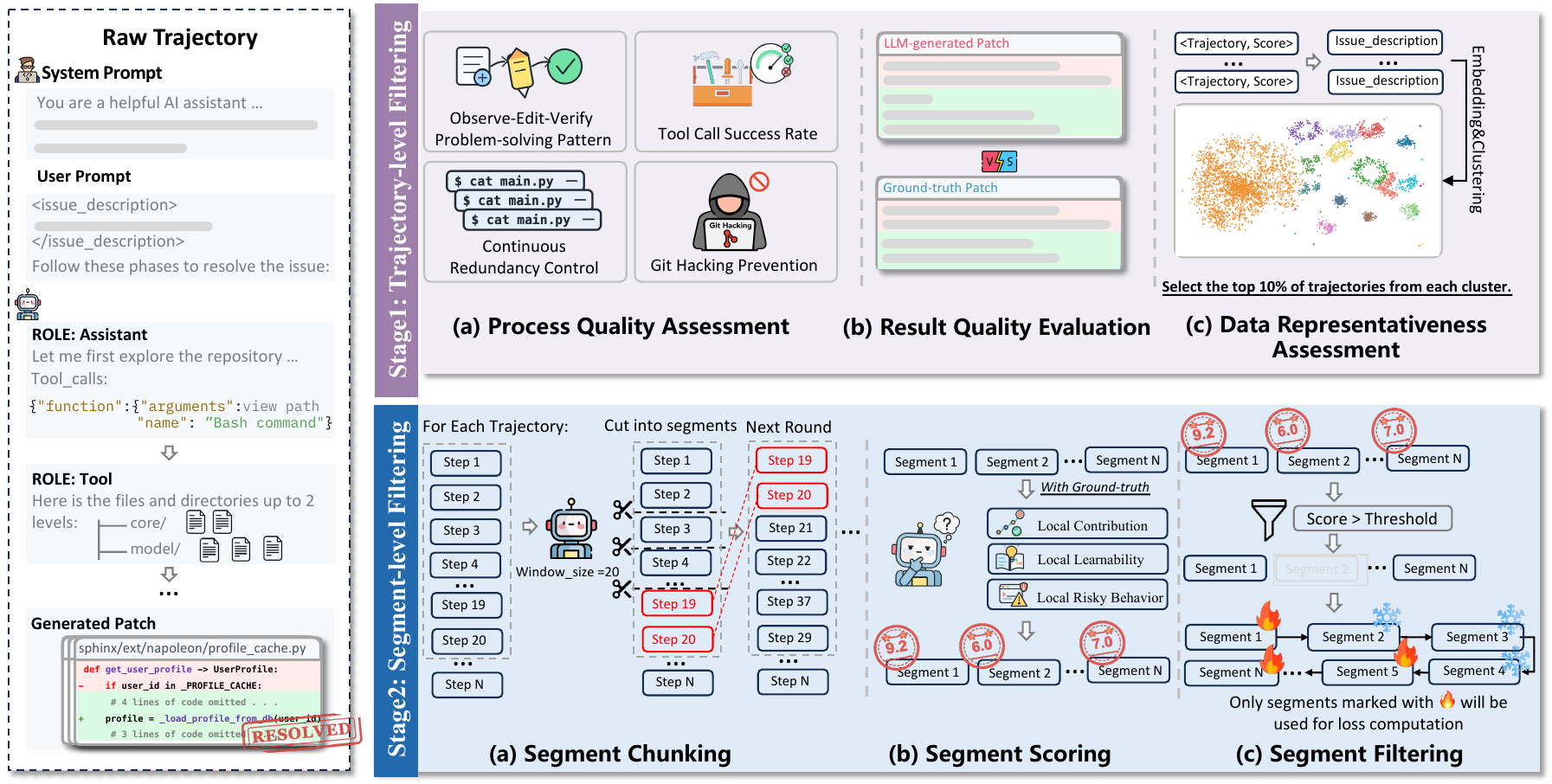}
    \caption{Overview of \sweprime, a multi-granularity, two-stage SFT data selection framework.}
    \label{fig:pipeline}
\end{figure*}

\section{Related Work}

\subsection{Software Issue Resolution}

Software issue resolution requires coding agents to generate repair patches for repository-level issues described in natural language. Addressing such issues involves understanding repository context and iteratively locating relevant code, implementing changes, and validating that the resulting patch resolves the issue without regressing existing functionality~\cite{jimenez2024swebench,yang2024sweagent,tao2024magis,zheng2025humanevo,zhang2024autocoderover,xia2025agentless,jiang2026phoenixrepair}.

\subsection{Training Data for Coding Agents}

SFT of coding agents relies on long-horizon interaction trajectories grounded in code repositories, tool feedback, and execution signals. Accordingly, existing work has focused on scaling the construction of executable tasks and the collection of test-verified trajectories. Executable tasks are commonly constructed from real-world issues, synthetic bugs, or code commits~\cite{yang2025swesmith,jain2025r2egym}. SWE-Gym, for example, constructs executable environments from GitHub issues and collects test-verified trajectories~\cite{pan2024swegym}. Successful trajectories are identified and retained through executable validation, soft verification signals, or filters based on execution outcomes and trajectory attributes~\cite{shen2026sera,song2026swemaster,liang2026swenext}. For instance, SWE-Factory automates environment construction and task verification across multiple programming languages~\cite{guo2025swefactory}. Among these efforts, SWE-Lego applies additional filtering after data construction by excluding steps with failed tool calls from the training loss~\cite{tao2026swelego}.

However, task success and explicit error signals alone are insufficient for identifying high-quality supervision. At the trajectory level, successful demonstrations may still exhibit unreliable processes or unfocused patches; at the segment level, behaviors without explicit errors may still be redundant, irrelevant, or risky~\cite{xiong2024watch,chen2025atlas,chen2025step,wang2025steca,he2026webstar}. To bridge these gaps, \sweprime applies quality-aware selection at both the trajectory and semantic-segment levels, shifting the focus from task success alone to the supervision value of complete trajectories and the behaviors within them.

\section{\sweprime}

\subsection{Overview}

As illustrated in Figure~\ref{fig:pipeline}, \sweprime is a multi-granularity, two-stage method that progressively selects SFT supervision from successful software issue-resolution trajectories. Stage~1 selects high-quality and representative trajectories, while Stage~2 identifies high-value semantic segments within them. During SFT, complete trajectories are retained as context, while only selected segments contribute to the loss.

\subsection{Stage 1: Trajectory-Level Selection}

Stage~1 performs trajectory-level selection by jointly considering three dimensions: process quality, result quality, and data representativeness.

\paragraph{Process Quality Assessment.}

Our process-quality assessment examines whether a successful trajectory provides a sound problem-solving demonstration. We operationalize it using four complementary signals observable from interaction logs: workflow grounding through observe-edit-verify, execution reliability through tool-call success, interaction efficiency through continuous-redundancy detection, and behavioral legitimacy through Git-hacking detection. Together, these signals cover complementary layers of process quality, connecting the validity of individual actions with the coherence of the overall problem-solving trajectory to assess its suitability for SFT supervision.

\noindent\textit{Observe-Edit-Verify Problem-solving Pattern.} An observe-edit-verify workflow grounds code changes in repository evidence and execution feedback, helping models learn systematic issue-resolution behaviors during SFT~\cite{chen2026beyondfinalcode}. We set $S_{\mathrm{workflow}}=1$ if a trajectory inspects the repository before its first edit and validates the patch through a test or check after its final edit, and 0 otherwise.

\noindent\textit{Tool Call Success Rate.} Frequent tool failures can introduce unreliable tool-use patterns into SFT supervision~\cite{yang2025toolmind}. We calculate the tool-call success rate for each trajectory and define its tool reliability score $S_{\mathrm{tool}}$ as the percentile rank of this rate within the trajectory pool. A higher score indicates more reliable tool use relative to other trajectories.

\noindent\textit{Continuous Redundancy Control.} During long-horizon interactions, agents may lose track of prior actions and repeat the same tool call with unchanged arguments in adjacent steps, potentially teaching models inefficient interaction patterns during SFT~\cite{xiao2025agentdiet}. We compare the tool names and arguments of adjacent calls and set $S_{\mathrm{redundancy}}=0$ if any adjacent pair matches and 1 otherwise.

\noindent\textit{Git Hacking Prevention.} Git hacking occurs when an agent uses future repair information from repository history to guide code changes, introducing leakage-based shortcuts into SFT supervision~\cite{song2026swemaster,tao2026swelego}. To detect such leakage, \sweprime examines whether Git-history operations reveal reference-patch content before the agent modifies the code. We set $S_{\mathrm{git}}=0$ when such leakage is detected and 1 otherwise.

\paragraph{Result Quality Evaluation.}

A high-quality patch should resolve the issue with minimal yet sufficient changes, avoiding unnecessary modifications that increase the risk of regressions~\cite{mockus2000predicting,kamei2013large,chen2025patchquality}. Training on overly broad patches may also encourage the model to over-edit during SFT. We therefore use the reference patch only to estimate an appropriate modification scope. We assess this scope at both file and line levels, capturing the breadth and size of a patch, respectively.

Let $F_{\mathrm{model}}$ and $F_{\mathrm{gold}}$ denote the numbers of files modified by the generated and reference patches, respectively, and let $L_{\mathrm{model}}$ and $L_{\mathrm{gold}}$ denote their numbers of changed lines. We define the file- and line-level scope scores as

\[
S_{\mathrm{file}} = \min\left(\frac{F_{\mathrm{gold}}}{F_{\mathrm{model}}},1\right),
\qquad
S_{\mathrm{line}} = \min\left(\frac{L_{\mathrm{gold}}}{L_{\mathrm{model}}},1\right).
\]

Each score remains 1 when the generated patch does not exceed the corresponding reference scope and decreases as the modification scope expands. We multiply the two scores to obtain

\[
S_{\mathrm{result}} = S_{\mathrm{file}} \times S_{\mathrm{line}}.
\]

We use multiplication so that a patch receives a high result score only when both its file and line scopes are focused. After evaluating result quality, we combine it with the four process-quality scores:

\[
S_{\mathrm{traj}}
=
\frac{1}{5}
\left(
\begin{array}{c}
S_{\mathrm{workflow}} + S_{\mathrm{tool}} + S_{\mathrm{redundancy}} \\
{}+ S_{\mathrm{git}} + S_{\mathrm{result}}
\end{array}
\right).
\]

A higher $S_{\mathrm{traj}}$ indicates a trajectory with both a reliable problem-solving process and a focused repair. 

\paragraph{Data Representativeness Assessment.}

Selecting trajectories solely by their quality scores may bias the selection toward a few similar issue types, resulting in redundant supervision and limited coverage~\cite{liu2024deita,abbas2023semdedup,zhang2025harnessing}. We therefore apply HDBSCAN~\cite{mcinnes2017hdbscan} to group semantically similar issues, enabling cluster-wise selection that reduces redundancy and improves coverage across issue types.

Specifically, we first extract the issue description associated with each trajectory and encode it using Qwen3-Embedding-8B~\cite{zhang2025qwen3embedding}. We then apply HDBSCAN to the resulting embeddings to group semantically similar issues, treating each unclustered issue separately during selection. Finally, we rank trajectories by $S_{\mathrm{traj}}$ within each issue group and select the highest-ranked candidates under the target budget. The selected trajectories form the candidate set for Stage~2.

\subsection{Stage 2: Segment-Level Selection}

Stage~2 performs segment-level selection within the trajectories retained by Stage~1 through three components: Segment Chunking, Segment Scoring, and Segment Filtering.

\paragraph{Segment Chunking.}

Individual steps are too narrow in scope for behavior-level quality assessment. A single step often captures only an action or observation, without sufficient context to determine its intent and outcome. We therefore group consecutive steps that share a common intent into semantic segments, each capturing a coherent local behavior, such as tracing a symbol's usage across the repository.

\begin{figure}[h]
    \centering
    \includegraphics[width=\columnwidth]{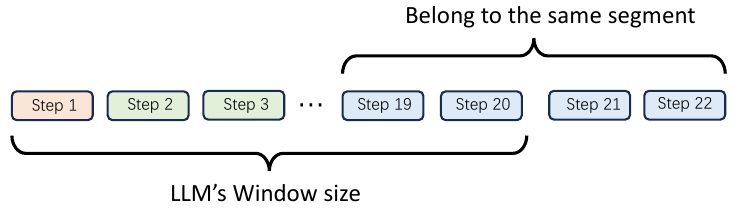}
    \caption{Semantic boundary truncation in fixed-size windows.}
    \label{fig:segment-chunking}
\end{figure}

Long trajectories may exceed the LLM's context window, so we process them incrementally using sliding windows~\cite{wang2026contextutilization}. However, fixed-size windows may split an ongoing behavior at their boundaries, leaving the final segment without sufficient subsequent context to determine its endpoint. As illustrated in Figure~\ref{fig:segment-chunking}, such boundary truncation can lead to incorrect segmentation and unreliable segment scoring.

To address this issue, \sweprime employs a boundary-aware chunking mechanism. After processing each window, \sweprime finalizes all segments except the last, which may be incomplete at the boundary. Then we carry this segment into the next window and reevaluates it together with subsequent steps to determine its semantic boundary. Additionally, to maintain consistent segmentation across windows, the LLM receives a compact summary of the task objective, established facts, and identified risks from preceding windows. Together, these mechanisms produce semantically coherent segments for subsequent quality assessment.

\paragraph{Segment Scoring.}

After obtaining semantically coherent segments, \sweprime further evaluates whether each behavior is worth using as SFT supervision~\cite{song2024trialerror,xiong2024watch,wang2025steca}. We define segment value according to three criteria: contribution to the final fix, local learnability, and behavioral risk.

\begin{itemize}
    \item \textbf{Contribution to the Final Fix.} A segment contributes to the final fix when it advances issue localization, context gathering, modification, verification, or error recovery.
    \item \textbf{Local Learnability.} We assess whether the behavior provides a clear and learnable evidence--action--feedback pattern based on the information available at that point.
    \item \textbf{Behavioral Risk.} Behavioral risk captures undesirable behaviors such as irrelevant exploration, uninformative failures, unsafe operations, and hardcoded solutions.
\end{itemize}

Guided by these three criteria, the LLM evaluates each segment using the issue description, trajectory context, and reference patch. It then assigns each segment $s_{i,j}$ a quality score $q_{i,j}\in[0,10]$ together with a rationale.

\paragraph{Segment Filtering.}

We construct segment-level training targets by assigning a binary selection mask to each segment according to its quality score. Specifically, $m_{i,j}=1$ if $q_{i,j}\geq\delta$, indicating that $s_{i,j}$ is selected as a learning target, and $m_{i,j}=0$ otherwise.

All segments are retained as context for subsequent behaviors, while the training loss is applied only to assistant-response tokens in selected segments~\cite{chen2025atlas,chen2025step,he2026webstar}. Accordingly, let $y_i$ denote the complete token sequence of trajectory $i$, and let $\mathcal{A}(s_{i,j})$ denote the assistant-response token positions in segment $s_{i,j}$. The selective SFT objective is

\[
\begin{array}{rl}
\mathcal{L}_{\mathrm{SFT}}
& =
-\frac{1}{N_{\mathrm{sel}}}
\sum_i\sum_j m_{i,j}
\sum_{t\in\mathcal{A}(s_{i,j})} \\[-1pt]
& \quad
\log p_\theta(y_{i,t}\mid y_{i,<t}), \\[2pt]
N_{\mathrm{sel}}
& =
\sum_i\sum_j m_{i,j}
\left|\mathcal{A}(s_{i,j})\right|.
\end{array}
\]

Here, $N_{\mathrm{sel}}$ denotes the total number of assistant-response tokens in selected segments and normalizes the objective into an average token-level loss. The mask $m_{i,j}$ prevents unselected segments from contributing to the loss, while their tokens remain in $y_{i,<t}$ as context for subsequent predictions.

\begin{table*}[t]
    \centering
    \caption{Main results on SWE-Bench Verified and SWE-Bench Pro. Rel. Imp. is relative to the raw model; Turns denotes average interaction turns.}
    \label{tab:main-results}
    \fontsize{7.6}{9.1}\selectfont
    \setlength{\tabcolsep}{1pt}
    \renewcommand{\tabularxcolumn}[1]{m{#1}}
    \begin{tabularx}{\textwidth}{>{\raggedright\arraybackslash}m{0.16\textwidth}>{\centering\arraybackslash}X!{\color{gray!75}\vrule width 0.6pt}*{2}{>{\centering\arraybackslash}X}!{\color{gray!75}\vrule width 0.6pt}>{\centering\arraybackslash}X!{\color{gray!75}\vrule width 0.6pt}*{4}{>{\centering\arraybackslash}X}!{\color{gray!75}\vrule width 0.6pt}*{2}{>{\centering\arraybackslash}X}!{\color{gray!75}\vrule width 0.6pt}>{\centering\arraybackslash}X}
        \toprule
        \multirow{2}{*}{\textbf{Method}} & \multirow{2}{*}{\textbf{Data Ratio}} & \multicolumn{3}{!{\color{gray!75}\vrule width 0.6pt}c!{\color{gray!75}\vrule width 0.6pt}}{\textbf{SWE-Bench Verified}} & \multicolumn{7}{c}{\textbf{SWE-Bench Pro}} \\
        \cmidrule(lr){3-5}\cmidrule(lr){6-12}
        & & \textbf{Overall} & \textbf{Rel. Imp.} & \textbf{Turns} & \textbf{Python} & \textbf{JS} & \textbf{TS} & \textbf{Go} & \textbf{Overall} & \textbf{Rel. Imp.} & \textbf{Turns} \\
        \midrule
        \rowcolor{modelgray}
        \multicolumn{12}{c}{\textbf{Qwen3-30B-A3B-Instruct-2507}\quad\textit{30.5B total / 3.3B activated}} \\
        Raw Model & -- & 25.2 & -- & 47.1 & 12.78 & 4.55 & 12.06 & 2.50 & 8.21 & -- & 35.4 \\
        Resolved-Trajectory SFT & 100\% & 36.8 & +46.0\% & 65.2 & 25.19 & 6.82 & 21.99 & 7.86 & 16.83 & +105.0\% & 68.3 \\
Random-10\% SFT & 10\% & 26.6 & +5.6\% & 69.8 & 13.53 & 6.82 & 13.48 & 3.21 & 9.17 & +11.7\% & 74.6 \\
        \rowcolor{oursblue}
        \textbf{\sweprime} & 10\% & \textbf{39.6} & \textbf{+57.1\%} & 61.1 & \textbf{27.44} & \textbf{9.09} & \textbf{23.40} & \textbf{10.00} & \textbf{18.88} & \textbf{+130.0\%} & 62.2 \\
        \quad w/o Stage~2 & 10\% & 34.6 & +37.3\% & 64.7 & 25.56 & 6.82 & 21.99 & 8.21 & 17.10 & +108.3\% & 64.6 \\
        \addlinespace[2pt]
        \rowcolor{modelgray}
        \multicolumn{12}{c}{\textbf{GLM-4.7-Flash}\quad\textit{30B total / 3B activated}} \\
        Raw Model & -- & 40.4 & -- & 68.3 & 12.78 & 6.82 & 17.02 & 7.50 & 11.22 & -- & 35.9 \\
        Resolved-Trajectory SFT & 100\% & 41.4 & +2.5\% & 67.9 & 29.32 & 20.45 & 26.24 & 20.00 & 24.62 & +119.5\% & 79.7 \\
Random-10\% SFT & 10\% & 39.6 & -2.0\% & 63.4 & 12.41 & 6.82 & 16.31 & 7.14 & 10.81 & -3.7\% & 84.3 \\
        \rowcolor{oursblue}
        \textbf{\sweprime} & 10\% & \textbf{51.4} & \textbf{+27.2\%} & 56.0 & \textbf{33.08} & \textbf{29.55} & \textbf{27.66} & \textbf{20.36} & \textbf{26.95} & \textbf{+140.2\%} & 70.8 \\
        \quad w/o Stage~2 & 10\% & 43.6 & +7.9\% & 61.7 & 27.82 & 27.27 & 26.24 & 16.07 & 22.98 & +104.9\% & 76.4 \\
        \addlinespace[2pt]
        \rowcolor{modelgray}
        \multicolumn{12}{c}{\textbf{Qwen3-Coder-30B-A3B-Instruct}\quad\textit{30.5B total / 3.3B activated}} \\
        Raw Model & -- & 44.8 & -- & 50.1 & 39.10 & 25.00 & 30.50 & 20.71 & 29.55 & -- & 54.4 \\
        Resolved-Trajectory SFT & 100\% & 51.0 & +13.8\% & 75.6 & 39.10 & 29.55 & 31.91 & 23.21 & 31.05 & +5.1\% & 83.6 \\
Random-10\% SFT & 10\% & 41.2 & -8.0\% & 77.5 & 36.09 & 22.73 & 28.37 & 18.93 & 27.22 & -7.9\% & 86.1 \\
        \rowcolor{oursblue}
        \textbf{\sweprime} & 10\% & \textbf{53.2} & \textbf{+18.8\%} & 70.9 & \textbf{43.61} & \textbf{38.64} & \textbf{33.33} & \textbf{26.43} & \textbf{34.75} & \textbf{+17.6\%} & 70.9 \\
        \quad w/o Stage~2 & 10\% & 50.2 & +12.1\% & 74.2 & 41.73 & 31.82 & 30.50 & 20.71 & 30.92 & +4.6\% & 79.7 \\
        \bottomrule
    \end{tabularx}
\end{table*}

\section{Experimental Setup}

\subsection{Research Questions}

\noindent\textbf{RQ1 (Effectiveness):} How effective is \sweprime in improving the software issue resolution capabilities of LLMs?

\noindent\textbf{RQ2 (Hyperparameter Analysis):} How sensitive is \sweprime to its hyperparameters, and does the chosen configuration generalize across settings?

\noindent\textbf{RQ3 (Behavioral Analysis):} How are the performance gains of \sweprime reflected in model behavior?

\subsection{Training Data, Models, and Benchmarks}

\paragraph{Training Data.}

We use the SWE-rebench OpenHands Trajectories dataset released by Nebius,\footnote{\url{https://huggingface.co/datasets/nebius/SWE-rebench-openhands-trajectories}} which contains 67,074 trajectories generated by Qwen3-Coder-480B-A35B-Instruct with OpenHands~\cite{wang2025openhands} on SWE-rebench issues~\cite{badertdinov2025swerebench}. Among them, 32,161 trajectories successfully resolve their corresponding issues. Since our study focuses on supervision quality among successful trajectories, we use these resolved trajectories as the initial candidate pool.

\paragraph{Base Models and Evaluation Benchmarks.}

We evaluate \sweprime on three base models: GLM-4.7-Flash\footnote{\url{https://huggingface.co/zai-org/GLM-4.7-Flash}}~\cite{zeng2025glm45}, Qwen3-30B-A3B-Instruct-2507\footnote{\url{https://huggingface.co/Qwen/Qwen3-30B-A3B-Instruct-2507}}~\cite{yang2025qwen3}, and Qwen3-Coder-30B-A3B-Instruct\footnote{\url{https://huggingface.co/Qwen/Qwen3-Coder-30B-A3B-Instruct}}~\cite{yang2025qwen3}. All models are evaluated on two software issue-resolution benchmarks: SWE-Bench Verified and SWE-Bench Pro. SWE-Bench Verified contains 500 human-validated tasks,\footnote{\url{https://openai.com/index/introducing-swe-bench-verified/}} while we use the 731-task public split of SWE-Bench Pro~\cite{deng2025swebenchpro}.

\subsection{Baselines}

\begin{itemize}
    \item \textbf{Raw Model.} We evaluate the original model without trajectory SFT to measure its pre-SFT performance.
    \item \textbf{Resolved-Trajectory SFT.} Following prior work that retains successful trajectories based on task-level execution outcomes~\cite{pan2024swegym,jain2025r2egym,guo2025swefactory}, we train on all 32,161 resolved trajectories without further quality filtering.
    \item \textbf{Random-10\%.} To assess the value of targeted selection under the same trajectory budget, we train on a random 10\% subset of complete resolved trajectories.
\end{itemize}

\subsection{Implementation Details}

\paragraph{Training Configuration.}

Following the training recipe released with SWE-rebench OpenHands Trajectories, we set the maximum sequence length to 131,072 and the batch size to 32. We use AdamW~\cite{loshchilov2019adamw} with a cosine learning-rate schedule and a peak learning rate of $4\times10^{-6}$. All settings optimize the cross-entropy objective, while \sweprime applies the loss only to assistant tokens in selected segments. Except for the training data and loss masks, all settings use the same training configuration.

\paragraph{Evaluation Setting.}

Each task is allowed up to 100 interaction turns with a maximum sequence length of 131,072. Sampling parameters follow each model's official defaults. Final patches are validated in the execution environment provided by each benchmark.

\section{Results}

\subsection{RQ1: Effectiveness}

\noindent\textbf{Overall Effectiveness.} \sweprime consistently achieves the highest Resolved Rate across all three models and both benchmarks, as shown in Table~\ref{tab:main-results}. The Rel. Imp. columns use the corresponding raw models as the reference: \sweprime yields relative improvements of 18.8\%--57.1\% on SWE-Bench Verified and 17.6\%--140.2\% on SWE-Bench Pro. These consistent gains across models with different initial capabilities show that \sweprime is effective across the evaluated model and benchmark settings.

\noindent\textbf{Data Efficiency.} Although \sweprime retains only 10\% of the resolved trajectories, it outperforms Resolved-Trajectory SFT across all evaluated settings. Relative to Resolved-Trajectory SFT, the gains vary across models and reach as high as 24.2\% on SWE-Bench Verified and 12.2\% on SWE-Bench Pro. These results highlight the importance of supervision quality: retaining more successful trajectories does not necessarily improve downstream performance, while a smaller, carefully selected subset can provide more effective supervision. \sweprime also substantially outperforms Random-10\%, further demonstrating the benefit of quality-aware selection under the same trajectory budget.

\noindent\textbf{Contribution of Segment-Level Selection.} To assess the contribution of segment-level selection, we compare \sweprime with its Stage~1-only variant, which performs trajectory-level selection without segment-level selection. Compared with this variant, \sweprime consistently achieves higher Resolved Rates across all evaluated settings, suggesting that trajectory-level filtering alone does not fully address low-value behaviors within otherwise high-quality trajectories. Segment-level selection therefore provides complementary benefits by focusing supervision on more valuable behaviors.

\noindent\textbf{Generalization Across Programming Languages.} To examine whether the effectiveness of \sweprime generalizes across programming languages, we compare its performance on the language-specific subsets of SWE-Bench Pro. Across all three base models, \sweprime achieves the best results on Python, JavaScript, TypeScript, and Go. This consistency aligns with the goal of cluster-based representative selection to reduce redundancy while retaining diverse issue types and broad supervision coverage in the selected training subset.

\subsection{RQ2: Hyperparameter Analysis}
\begin{figure}[t]
    \centering
    \includegraphics{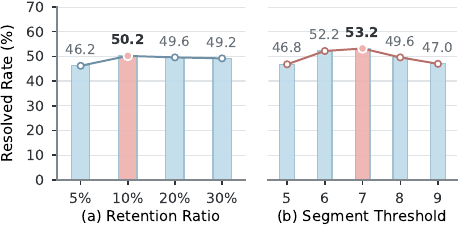}
    \caption{Hyperparameter sensitivity of Qwen3-Coder-30B-A3B-Instruct on SWE-Bench Verified.}
    \label{fig:hyperparameter-sensitivity}
\end{figure}

\begin{figure*}[t]
    \centering
    \includegraphics[width=\textwidth]{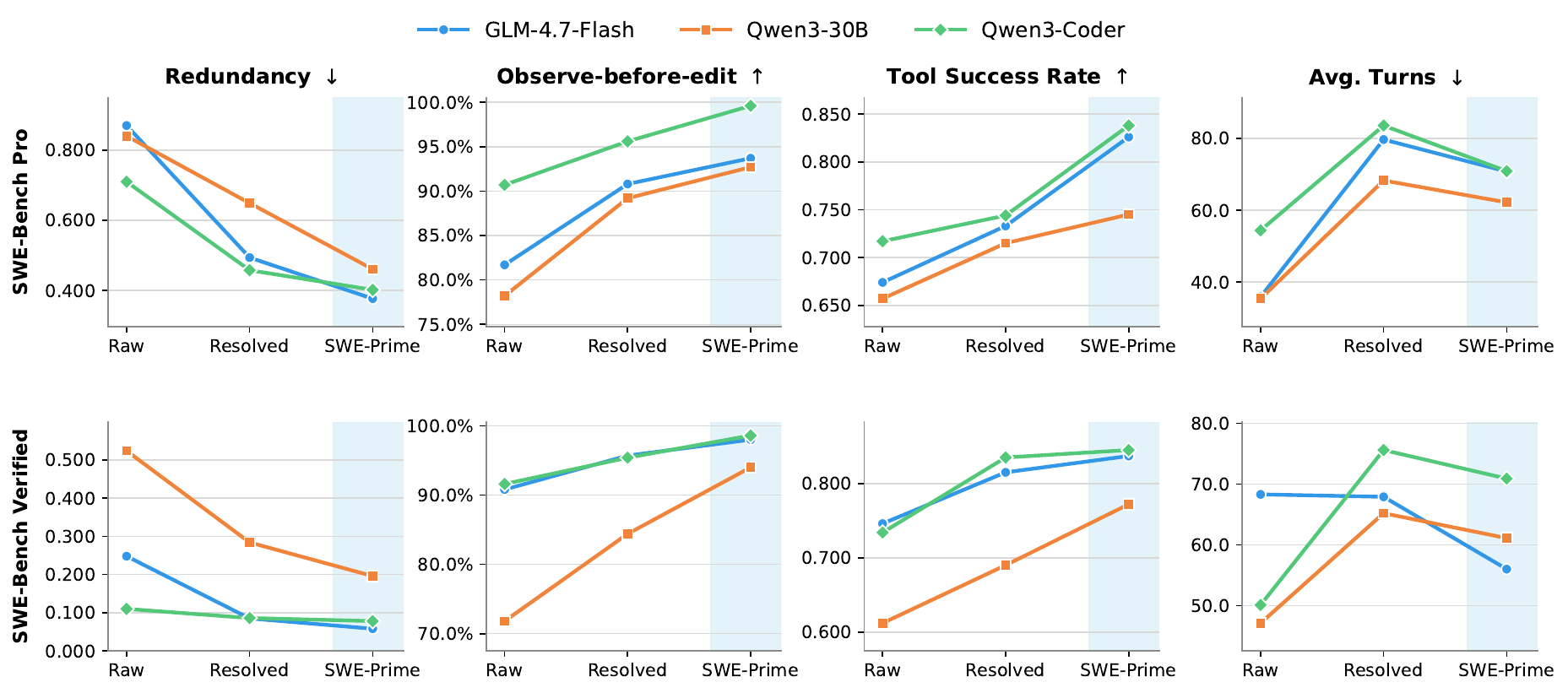}
    \caption{Behavioral comparison of the Raw Model, Resolved-Trajectory SFT, and \sweprime. Arrows indicate preferred directions.}
    \label{fig:behavioral-analysis}
\end{figure*}

\noindent\textbf{Analysis Protocol.} Because the proportion of high-quality supervision in a trajectory pool is not known a priori, we first determine the selection configuration of \sweprime before conducting the main evaluation in RQ1. We examine two hyperparameters. The Stage~1 trajectory retention ratio controls the number of retained trajectories, whereas the Stage~2 segment score threshold governs segment inclusion in the training loss. We conduct this exploratory analysis with Qwen3-Coder-30B-A3B-Instruct on SWE-Bench Verified, first examining the trajectory retention ratio and then fixing it while analyzing the segment score threshold. \textbf{Once selected, the resulting configuration is frozen and applied to all experiments reported in RQ1.}

\noindent\textbf{Trajectory Retention Ratio.} To isolate the effect of trajectory-level selection, we disable Stage~2 and perform standard SFT directly on the complete trajectories retained by Stage~1, with all assistant tokens contributing to the training loss. We vary the retention ratio among 5\%, 10\%, 20\%, and 30\%. As shown in Figure~\ref{fig:hyperparameter-sensitivity}(a), the Resolved Rate increases from 46.2\% at 5\% retention to a peak of 50.2\% at 10\%, and then slightly decreases to 49.6\% and 49.2\% at 20\% and 30\%, respectively. This pattern suggests that retaining too few trajectories may limit data coverage, while including additional trajectories beyond 10\% provides no further benefit and may reintroduce lower-quality or redundant supervision. We therefore select a trajectory retention ratio of 10\%.

\noindent\textbf{Segment Score Threshold.} After selecting the 10\% retention ratio, we fix it and examine the effect of the Stage~2 segment score threshold. All segments remain in their original trajectories to preserve context, while only segments meeting the threshold contribute to the training loss. As shown in Figure~\ref{fig:hyperparameter-sensitivity}(b), increasing the threshold from 5 to 7 improves the Resolved Rate from 46.8\% to 53.2\%. Further increasing the threshold to 8 and 9 reduces the performance to 49.6\% and 47.0\%, respectively. A lower threshold may retain more low-value supervision, whereas an overly restrictive threshold may exclude useful segments. We therefore set the segment score threshold to 7.

\noindent\textbf{Generalization of the Selected Configuration.} Based on the above analysis, we select a 10\% trajectory retention ratio and a segment score threshold of 7, and apply this configuration unchanged throughout RQ1. As shown in Table~\ref{tab:main-results}, \sweprime consistently improves performance across all three base models and both benchmarks, suggesting that the selected configuration remains effective beyond the model and benchmark used to determine it. Although we cannot know in advance how many trajectories in a given pool provide high-quality supervision, this consistency suggests that the selected configuration achieves an effective balance between supervision quality and coverage across the evaluated settings.

\subsection{RQ3: Behavioral Analysis}

\noindent\textbf{Analysis Protocol.} To examine how the performance gains of \sweprime are reflected in model behavior, we analyze interaction trajectories from two complementary perspectives: process quality and interaction efficiency. RQ1 establishes the improvements in Resolved Rate, but this outcome metric does not reveal how models arrive at their solutions. We therefore compare the trajectories generated by the Raw Model, Resolved-Trajectory SFT, and \sweprime on the evaluation tasks. We focus primarily on Resolved-Trajectory SFT, which uses the same candidate pool and training configuration, and treat the Raw Model as a pre-SFT reference. As shown in Figure~\ref{fig:behavioral-analysis}, process quality is measured by redundancy, observe-before-edit rate, and tool success rate, while interaction efficiency is measured by the average number of interaction turns.

\noindent\textbf{Process Quality.} Models trained with \sweprime consistently improve all three process-quality metrics across the evaluated models and benchmarks. Compared with Resolved-Trajectory SFT, \sweprime reduces redundancy by up to 0.188, increases the observe-before-edit rate by up to 9.6 percentage points, and improves the tool success rate by up to 0.094. These changes correspond to fewer repeated actions that provide no new information, more frequent context gathering before code modification, and more reliable tool use. Their consistency across metrics suggests that the performance gains of \sweprime are accompanied by more reliable and structured problem-solving behavior.

\noindent\textbf{Interaction Efficiency.} As shown in Figure~\ref{fig:behavioral-analysis}, \sweprime reduces the average number of interaction turns relative to Resolved-Trajectory SFT across all six evaluation settings, with reductions ranging from 4.1 to 12.7 turns. Because these reductions are consistently accompanied by higher Resolved Rates, they are unlikely to result from premature termination. Instead, this pattern reflects the role of segment-level selection: by preventing low-value behaviors from contributing to the SFT loss, \sweprime helps models learn more focused and efficient issue-resolution patterns.

\section{Conclusion}

This work studies how to select high-quality SFT supervision from successful coding-agent trajectories, as task success alone cannot determine whether the recorded process is suitable for learning. We introduce \sweprime, a multi-granularity, two-stage method that selects a high-quality and representative trajectory subset based on process quality, result quality, and data representativeness, then identifies valuable semantic segments based on their contribution, learnability, and potential risks. On SWE-Bench Pro and SWE-Bench Verified, models trained on only 10\% of the trajectories selected by \sweprime outperform those trained on the entire resolved-trajectory pool, with relative performance gains of up to 12.2\% and 24.2\%, respectively. These results suggest that effective coding-agent SFT depends not only on successful task outcomes, but also on the quality, representativeness, and learnability of the resulting supervision.